\documentclass[a4paper,11pt]{article}
\usepackage{pos}

\usepackage{xifthen}
\usepackage{dsfont}
\usepackage[titletoc]{appendix}
\usepackage{booktabs}
\usepackage{units}
\usepackage{braket}

\newcommand{\gettitle}{}
\hypersetup{linkcolor=black
	colorlinks,
	linkcolor={red!75!black},
	citecolor={blue!75!black},
	urlcolor={blue!75!black},
	pdftitle={\gettitle},
	pdfauthor={Barata, Mehtar-Tani},
	pdfkeywords={Perturbative QCD} {Jet quenching},
	bookmarksopen=true,
	bookmarksopenlevel=2,
	bookmarksnumbered=true
}
\makeatletter
\newcommand\makebig[2]{%
  \@xp\newcommand\@xp*\csname#1\endcsname{\bBigg@{#2}}%
  \@xp\newcommand\@xp*\csname#1l\endcsname{\@xp\mathopen\csname#1\endcsname}%
  \@xp\newcommand\@xp*\csname#1r\endcsname{\@xp\mathclose\csname#1\endcsname}%
}
\makeatother
\makebig{biggg} {1.3}

\long\def\comment#1{ }

\newcommand{\nn}{\nonumber\\ }

\def\be{\begin{eqnarray*}}
\def\ee{\end{eqnarray*}}
\def\beq{\begin{eqnarray}}
\def\eeq{\end{eqnarray}}
\newcommand{\bea}{\beq \begin{aligned}}
\newcommand{\eea}{\end{aligned}\eeq}

\def\k{{\boldsymbol k}}

\def\0{{\boldsymbol 0}}
\def\l{{\boldsymbol l}}
\def\k{{\boldsymbol k}}

\DeclareMathOperator\erf{erf}

\def\and{ \quad\text{and}\quad}

\def\Re{\text{Re}}

\def\cT{{\cal T}}

\title{Energy loss effects in EECs at LO}

\author*[a]{João Barata}
\author[a,b]{Yacine Mehtar-Tani}

\affiliation[a]{Physics Department, Brookhaven National Laboratory, Upton, NY 11973, USA}

\affiliation[b]{RIKEN BNL Research Center, Brookhaven National Laboratory, Upton, NY 11973, USA}

\emailAdd{jlourenco@bnl.gov}
\emailAdd{mehtartani@bnl.gov}

\abstract{In recent years, there has been an effort towards establishing a more complete picture for jet substructure in the presence of the quark gluon plasma. Such a program requires not only a more detailed description of medium induced effects, but also the design of novel substructure observables. Very recently, it has been noticed that Energy Energy correlators (EECs) might provide one type of such observables. Although the full extent of their sensitivity to the medium has not been completely explored, they are capable to resolve the transverse structure of the jet. In particular, they are sensitive to the critical angle separating coherent and decoherent jet evolution in the medium. In this talk, we show for the first time the effects of medium induced radiative energy loss in EECs at leading order in the number of vacuum-like emissions. The calculation takes into account all order soft gluon emissions, in the large $N_c$ limit and neglecting subdominant interfering contributions.}

\FullConference{
 26-31 March 2023 \\
 Aschaffenburg, Germany\\}

\begin{document}
\maketitle

\section{Introduction}
In recent years, there has been a strong interest in applying jet substructure techniques to study medium induced modifications to jets. One of the most interesting ideas being currently pursued relates to measuring the so called critical coherence angle $\theta_c$, which determines the radiation pattern of jets in medium, see e.g.~\cite{Casalderrey-Solana:2012evi, Blaizot:2015lma}. As far as we are aware, two recent proposals were made to extract information related to this parameter: one resorts to measuring the angular distribution associated to groomed emissions~\cite{Caucal:2021cfb}, while the other uses the information related to energy flow correlations inside jets~\cite{Andres:2022ovj}. Although these observables have very different properties, it is interesting to note that while for the former the coherence angle effects enter mainly through the energy loss mechanism, in the latter it emerges due to the medium modified splitting kernel. Thus, our goal is to explore the sensitivity of the energy flow correlations to radiative energy loss.

\section{Energy Energy Correlators and medium modified cross-section}
The idea of understanding high energy scattering events using energy flow correlations
was proposed a long time ago~\cite{Basham:1978bw}. Recently, these ideas have been extended to study the structure of QCD jets in vacuum~\cite{Lee:2022ige} and in heavy ion collisions~\cite{Andres:2022ovj,Andres:2023xwr}. At leading order in the strong coupling constant and assuming an isotropic and homogenous medium, one can define the two point energy flow correlator (EEC) as~\cite{Andres:2023xwr}
\begin{align}~\label{eq:general_unquench}
	\frac{d\Sigma}{d\theta} &\equiv   \int_0^1 dz \, z(1-z) \frac{1}{\sigma_{qg}}\frac{d\sigma_{qg}}{d\theta dz}  \, ,
\end{align}
where we have focused on a jet originating from a single quark, $z$ denotes the energy sharing fraction for the outgoing gluon, $\theta$ is the angle between the outgoing states and we neglect the initial quark cross-section in what follows. In the vacuum and at leading accuracy, this object is fully determined by 
\begin{align}
  \frac{1}{\sigma_{qg}}\frac{d\sigma_{qg}^{\rm vac}}{d\theta dz}  = \frac{\alpha_s C_F}{\pi} P(z) \frac{1}{ \theta} \, ,
\end{align}
where the relevant splitting function is given by $P(z) = (1+(1-z)^2)/z$. In the presence of a dense QCD medium, the partonic cross-section gets corrections due to the interactions with the medium constituents. Although the general form of such a cross-section is known, it is a highly complex object for practical applications, see e.g.~\cite{Apolinario:2014csa}. Therefore, for phenomenology, different limits of this are typically used. Here we shall consider two limiting cases: 
\begin{enumerate}
  \item \textbf{Hard splitting}: It is natural to consider the scenario where the outgoing partons are very energetic, and thus they propagate close to their classical trajectories. In this limit, it has been shown that one can write~\cite{Dominguez:2019ges,Isaksen:2020npj}
  \begin{align}
    \frac{d\sigma_{qg}^{\rm med}}{d\theta dz}= \frac{d\sigma_{qg}^{\rm vac}}{d\theta dz}	\left(1+ F_{\rm med }(\theta,z)\right) \, , 
  \end{align}
  where the medium modifications are encapsulated in the $F_{\rm med }$ factor, which reads 
  \begin{align}\label{eq:F_med}
    F_{\rm med}= \frac{2}{t_f} \int_0^L dt\, \left\{\frac{\left(1-e^{-\frac{1}{4} \chi \hat q \theta^2 t^2 (L-t)} \right)}{t_f(\frac{1}{4} \chi \hat q 
    \theta^2 t^2)} \cos\left(\frac{t}{t_f}\right) e^{-\frac{1}{12} \hat q \theta^2 \zeta t^3}  - \sin\left(\frac{L-t}{t_f}\right) e^{-\frac{1}{12} \hat q \zeta \theta^2 (L-t)^3 }\right\}\, .
  \end{align}
  Here we have neglected subleading terms related to color interference, $N_c$ is the number of colors, $\hat q$ is the jet quenching parameter, $L$ the medium length and $t_f=\frac{2}{z(1-z) p_t\theta^2}$, with $p_t$ the total jet energy. We note that since the outgoing states are very energetic, this leads to the kinematical constraint $\min(z,1-z)p_t > \frac{1}{2}\, \hat q L^2$~\cite{Altinoluk:2015gia}. We have also introduced the variables $\zeta = 	\left( 1+z^2+\frac{2z}{N_c^2-1}\right) $ and $\chi = (1-2z+3z^2)$. In Fig.~\ref{fig:1} (left), we plot the integrand of Eq.~\ref{eq:general_unquench} using Eq.~\ref{eq:F_med}, as a function of the gluon energy fraction for several values of $\theta$. We observe that for angles close to $\theta_c\equiv \frac{2}{\sqrt{\hat q L^3}}$, the distribution has a flat dependence in $z$, but at large angles it is dominated by very asymmetrical topologies, with the gluon carrying a small amount of energy.

  \begin{figure}[h]
    \centering
    \includegraphics[width=.45\textwidth]{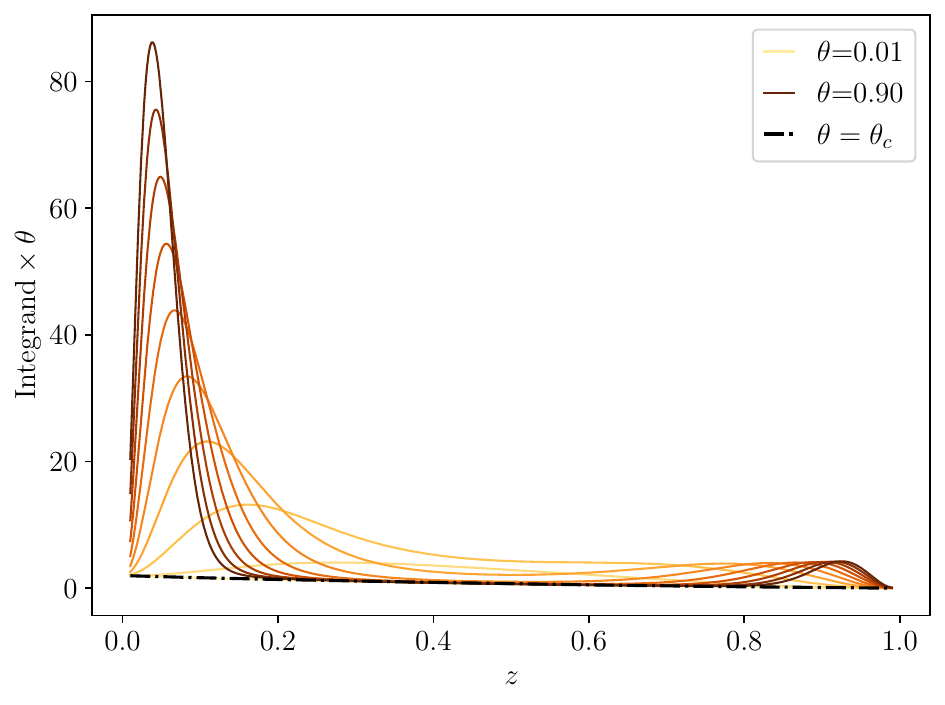}
    \includegraphics[width=.45\textwidth]{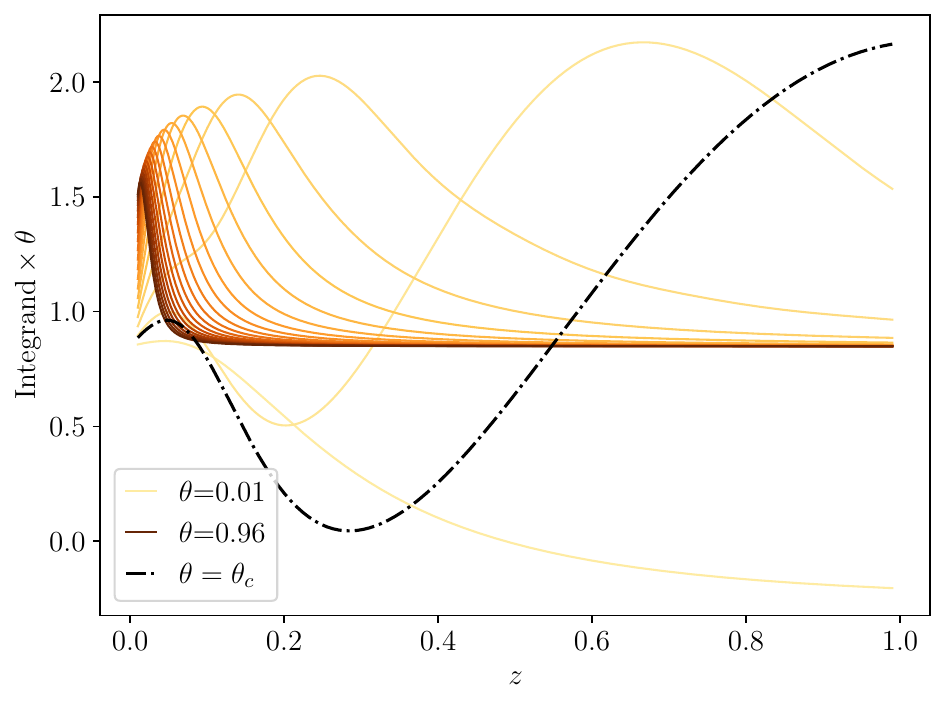}
    \caption{Integrand of the in-medium EEC: hard branching (left) and soft splitting (right). Here we used $p_t=100$ GeV, $\hat q =0.2\,{\rm GeV}^3$ and $L=6$ fm. As a result $\theta_c=0.047$.}
    \label{fig:1}
  \end{figure}

  \item   \textbf{Soft gluon splitting}: In  opposition to the previous case, we also consider the exact soft gluon limit, i.e $z\ll1$. In this case, the medium modified cross-section is given by
  \begin{align}
    \frac{d\sigma_{qg}^{\rm med}}{d\theta dz} =  (2\pi)^2 \frac{\omega dI}{d\omega d^2\k } \frac{zp_t^2 \theta}{2\pi}\Theta(z(1-z)E <\omega_c) \, ,
  \end{align}
where $2 \omega_c= \hat q L^2$, $\omega= z p_t$, $\theta \omega = |\k| $ and the single gluon spectrums reads (see e.g.~\cite{Barata:2021wuf})
  \begin{align}\label{eq:Inin_bdmps}
    (2\pi)^2 \omega\frac{dI}{d\omega d^2\k}&= \frac{4\alpha_s C_F}{z p_t } \Re \int_{0}^L ds  \, \frac{e^{\frac{(z p_t \theta)^2 \cT_{s}}{i-Q_{s}^2\cT_{s}}}}{Q_{s}^2\cT_{s} -i} -\frac{8\alpha_sC_F}{(z p_t \theta)^2} \, \Re \left(1-e^{-i (z p_t \theta)^2 \cT_L}\right)\, .
  \end{align}
Here $Q_{s}^2 = \hat{q} (L-s)$ and $\cT_s = \frac{\tan(\Omega s)}{2\omega \Omega}$, $\Omega=(1-i)/2 \sqrt{\hat q/(z p_t)}$. In Fig.~\ref{fig:1} (right) we again plot the EEC integrand for this limit of the in-medium cross-section. We observe that for angles close to $\theta_c$ the behavior differs significantly from the one seen in Fig.~\ref{fig:1} (left). However, at large angles, we observe that the system is again dominated by very asymmetrical configurations.

  
\end{enumerate}

\section{Jet energy loss}
The above results do not take into account the possibility of transporting the jet energy down to the temperature scale ($T$) by the emission of multiple gluons, which are thought to dominate the energy loss mechanism in matter, see e.g.~\cite{Blaizot:2013hx}. To include this effect here we use the quenching weight approximation~\cite{Baier:2001yt}, in the limit of two color sources being produced in the medium~\cite{Mehtar-Tani:2017ypq}. This allows us to model the jet energy loss via the relation ($t_c= 2/(\hat q  \theta^2)^{\frac{1}{3}}$)~\cite{Mehtar-Tani:2021fud}
\begin{align}
	\frac{d\Sigma(p_t)}{ d\theta} &=   \int_0^1 dz \, z(1-z)  \frac{1}{\sigma_{qg}}\frac{d\sigma_{qg}}{d\theta dz} \nn 
	&\times Q_q [\Theta(t_f>L) + \Theta(t_f<L) \Theta(\theta<\theta_c)  + Q_g \Theta(t_f<t_c) \Theta(\theta>\theta_c)  ] \,. 
\end{align}
Here $Q_{q(g)}$ denote the single quark (gluon) quenching factor, which can be 
computed, under some assumptions, as the modified Laplace transform over the single gluon spectrum
\begin{align}
   Q = \exp - \left\{ \int d\omega \int \frac{d^2\k}{(2\pi)^2} (2\pi)^2 \frac{dI}{d\omega d^2\k} \left(1- e^{-\eta \omega}\right) \right\}	\, ,
  \end{align}
with $Q$ standing for either the quark or gluon quenching weights. Here, we further model this object by assuming 1) all radiation between $T$ and the scale where gluon emission becomes abundant ($\omega_s=\alpha_s^2 C_A^2/\pi^2 \omega_c$) is lost to the medium 2) gluon radiation above this limit is only lost if emitted at angles larger than $2R$, with $R$ the jet radius. Then, one can write that\footnote{Here: $\eta = \frac{n}{p_t}$, $\alpha \equiv \frac{\chi}{\eta^2} = \left(\frac{2p_t R}{\sqrt{\hat q L}n}\right)^2	$ and $n\sim 6$ is the spectral index of the jet spectra.}
\begin{align}
  Q(p_t,R) &= \exp \Bigg\{- \Bigg(\frac{2\alpha_s C_i}{\pi}  \Bigg[ \sqrt{\frac{2\omega_c}{T}} (1-e^{-\eta T}) -  \sqrt{\frac{2\omega_c}{\omega_s}} (1-e^{-\eta \omega_s}) \nn 
  &+ \sqrt{2\eta \pi \omega_c} (\erf(\sqrt{\omega_s \eta}) -\erf(\sqrt{T \eta})  )\Bigg] \Bigg) \nn 
  &- \Bigg( \frac{\alpha_s C_i}{\pi} \sqrt{2\omega_c \eta} \int_{\eta \omega_s}^\infty dx\frac{1}{x^{\frac{3}{2}}} \left(1- e^{-x}\right) \left(e^{-x^2 \alpha}-\alpha x^2 \Gamma_0(\alpha x^2)\right)\Bigg) \Bigg\}	
\end{align}

\begin{figure}[h]
  \centering
  \includegraphics[width=.45\textwidth]{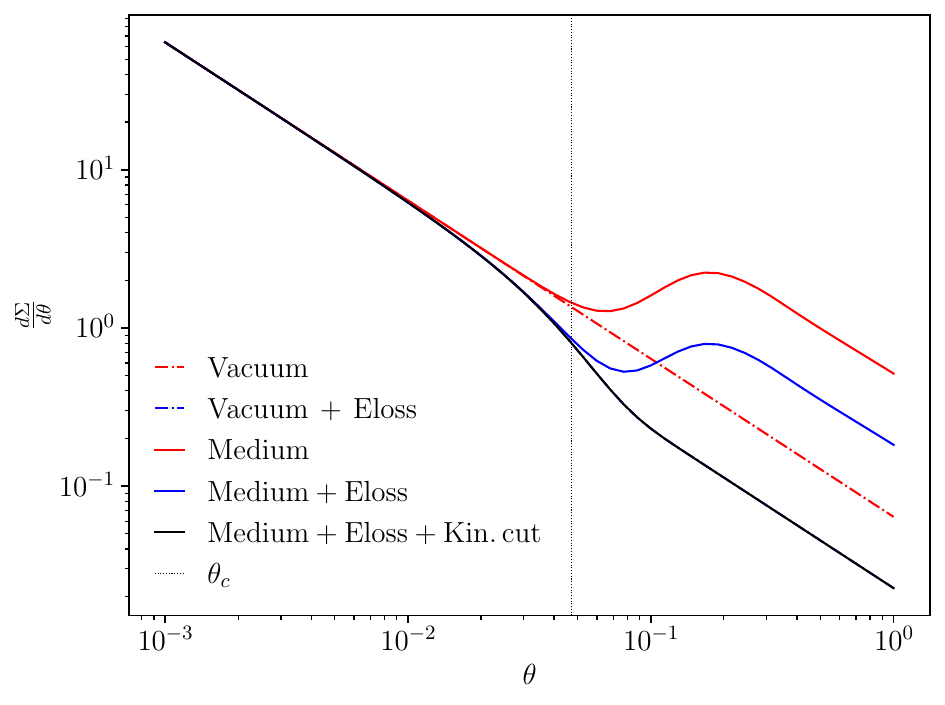}
  \includegraphics[width=.45\textwidth]{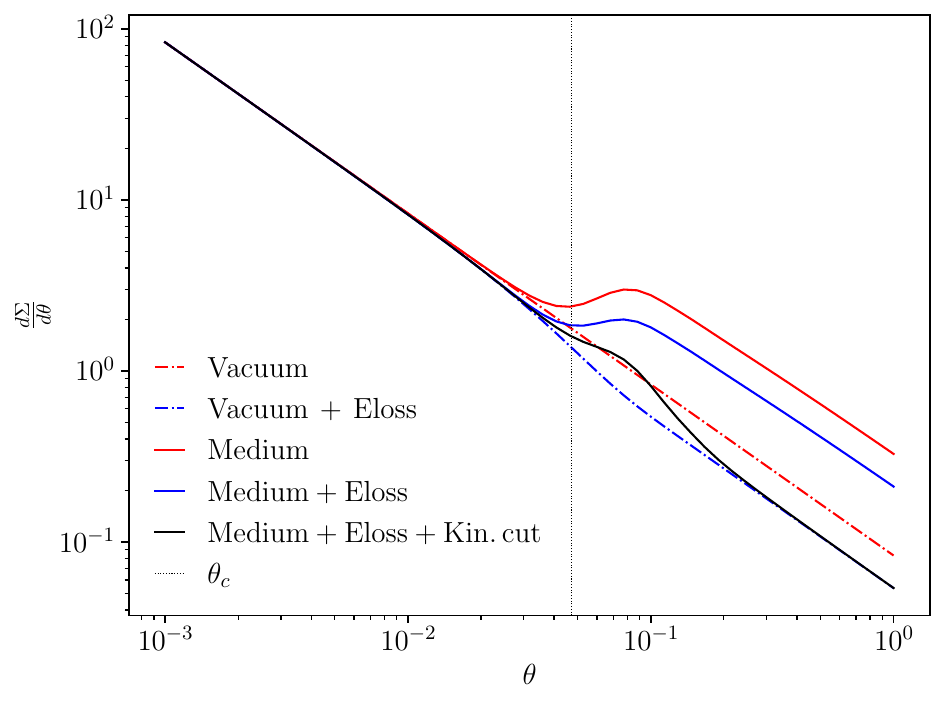}
  \caption{EEC distribution including energy loss using $p_t=100$ ($250$) GeV (right), $R=0.3$, $T=0.3$ GeV and the same parameters as in the previous plots. Dashed lines denote the vacuum results, solid red and blue curves correspond to including the medium modified cross-section. The vacuum lines including energy loss, convolute the vacuum kernel with the energy loss model.  The black line corresponds to the full result while imposing the constraint $\min(z,1-z)p_t>\omega_c$.}
  \label{fig:3}
\end{figure}

\begin{figure}[h]
  \centering
  \includegraphics[width=.45\textwidth]{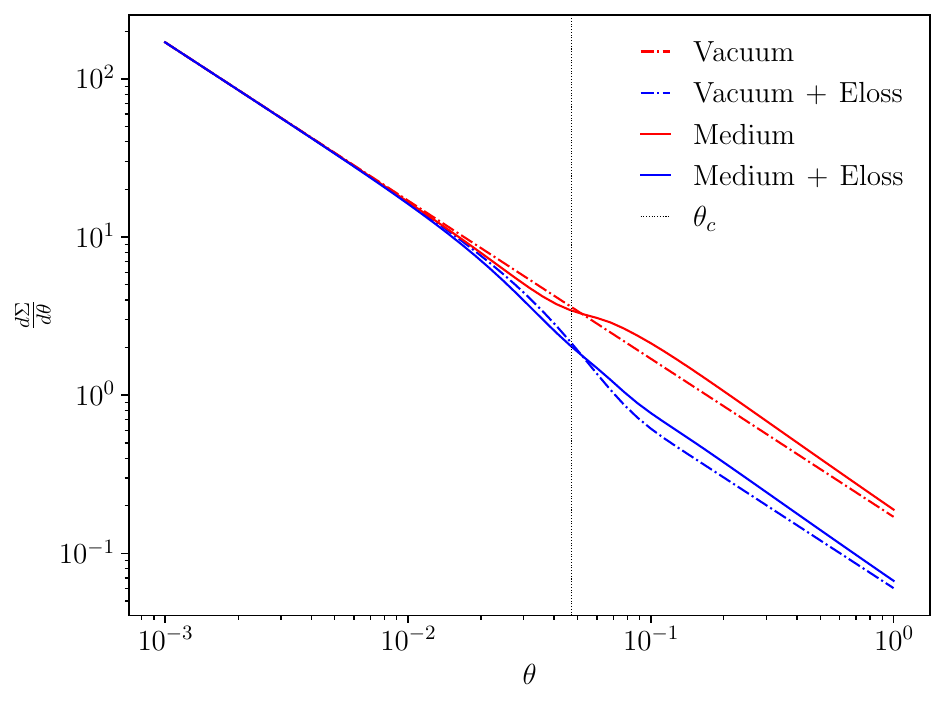}
  \includegraphics[width=.45\textwidth]{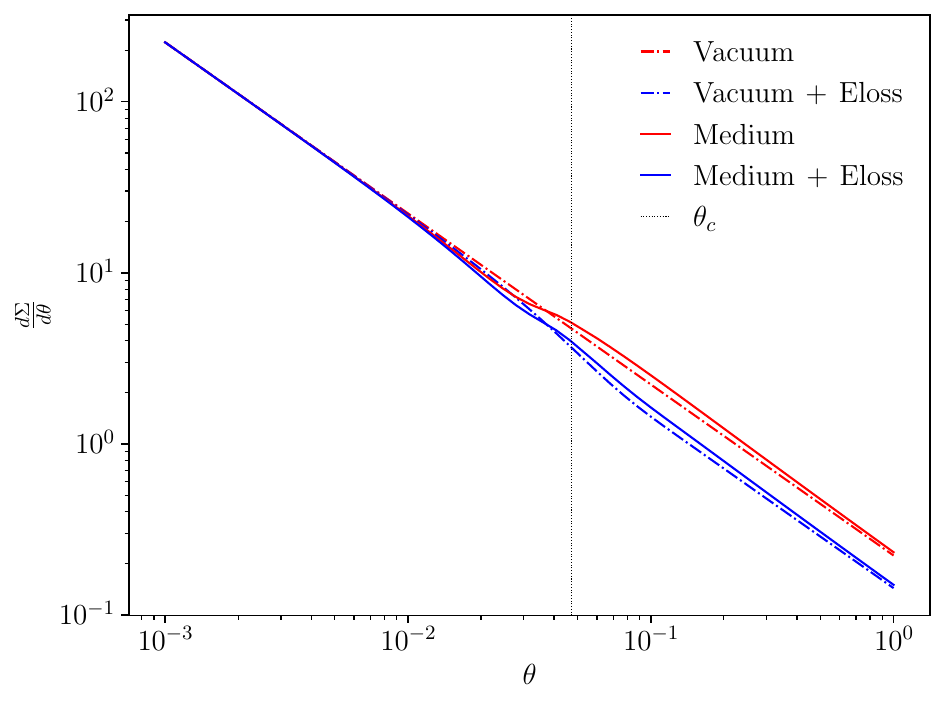}
  \caption{Same distributions as in Fig.~\ref{fig:3}, but for soft gluon branching.}
  \label{fig:4}
\end{figure}

In Figs.~\ref{fig:3} and \ref{fig:4}, we show the numerical results after taking into account radiative energy loss, for hard (Fig.~\ref{fig:3}) and soft splittings (Fig.~\ref{fig:4}). For the hard branchings, we observe that the energy loss does not change the shape of the distribution, unless a kinematical cut is imposed. In this case, energy loss leads to a dip with respect to the vacuum at large angles. For soft splittings, we observe a qualitatively similar behavior, despite the fact that this limit is more sensitive to the energy loss. We note that in this limit the validity of the description is stretched by integrating over the full $z$ domain.

\section{Conclusion and Discussion}
In this talk, we discussed the effects of energy loss in EECs measured inside jets produced in heavy ion collisions. These exploratory results indicate that energy loss effects might be important in the description of EECs, but a more careful treatment of the in-medium cross-sections is needed.

\section*{Acknowledgements}
J. B. is grateful to Alba Soto-Ontoso, Paul Caucal, Andrey Sadofyev and Robert Szafron for helpful discussions. We also are grateful to the authors of~\cite{Andres:2022ovj} for clarifications on their work. Y. M.-T. and J. B.'s work has been supported by the U.S. Department of Energy under Contract No.~DE-SC0012704. Y. M.-T. acknowledges support from the RHIC Physics Fellow Program of the RIKEN BNL Research Center.

\bibliographystyle{elsarticle-num}

\bibliography{Lib.bib}

\end{document}